\documentclass[preprint,aps]{revtex4}
\usepackage{graphicx}
\usepackage{dcolumn}
\usepackage{bm}

\begin{document}

\title{Sodium vacancy ordering and the co-existence of localized spins and itinerant charges in $Na_xCoO_2$}

\author{F. C. Chou$^{1,2}$, M. -W. Chu$^1$, G. J. Shu$^1$, F. T. Huang$^{1,3,4}$, Woei Wu Pai$^1$, H. S. Sheu$^2$, T. Imai$^5$, F. L. Ning$^5$, \& Patrick A. Lee$^6$}

\affiliation
{$^1$Center for Condensed Matter Sciences, National Taiwan University, Taipei 10617, Taiwan,\\
 $^2$National Synchrotron Radiation Research Center, HsinChu 30076,Taiwan,\\
 $^3$Taiwan International Graduate Program, Academia Sinica, Taipei 115, Taiwan,\\
 $^4$Department of Chemistry, National Taiwan University, Taipei 10617, Taiwan,\\
 $^5$Department of Physics and Astronomy, McMaster University, Hamilton, ON L8S4M1, Canada,\\
 $^6$Department of Physics, Massachusetts Institute of Technology, Cambridge, MA 02139\\}

\begin{abstract}
The sodium cobaltate family ($Na_xCoO_2$) is unique among transition metal oxides because the Co sits on a triangular lattice and its valence can be tuned over a wide range by varying the Na concentration $x$.  Up to now detailed modeling of the rich phenomenology (which ranges from unconventional superconductivity to enhanced thermopower) has been hampered by the difficulty of controlling pure phases. We discovered that certain Na concentrations are specially stable and are associated with superlattice ordering of the Na clusters.  This leads naturally to a picture of co-existence of localized spins and itinerant charge carriers.  For $x = 0.84$ we found a remarkably small Fermi energy of 87~K.  Our picture brings coherence to a variety of measurements ranging from
optical to thermal transport.  Our results also allow us to take the first step towards modeling the mysterious ``Curie-Weiss'' metal state at $x = 0.71$.  We suggest the local moments may form a
quantum
spin liquid state and we propose experimental test of our hypothesis.
\end{abstract}

\maketitle

Sodium cobaltate ($Na_xCoO_2$) has
received a great deal of attention from the condensed matter physics
and materials science communities recently because the cobalt
valence can be tuned over a great range by varying the Na
concentration, leading to unusual properties all the way from
unconventional superconductivity  when hydrated\cite{Takada} at $
x \approx 0.3$ to a novel ``Curie-Weiss metal''\cite{Foo} for $
x \approx 0.7$ to enhanced thermopower\cite{Lee} for $x \approx
0.85$.  Here we concentrate on $x > 0.5$ where unusual
magnetism\cite{Bayrakci05} has been observed and the ordering of the
Na ions is thought to play an important role.\cite{Mukhamedshin,Ning,Zandbergen,Meng}  Through
electrochemical de-intercalation and a careful characterization of
the Na concentration on high purity single crystals, we determine
that there is a special island of stability at $x =
0.71\pm 0.01$. We also studied $x = 0.84\pm0.01$ which is the
limit of single phase stability for high temperature melt growth.
Using x-ray and electron diffraction, we determine that these two
phases exhibit special Na ordering patterns.  We discovered a
hexagonal $\sqrt{13}a \times \sqrt{13}a$ superstructure for
$x = 0.84 \approx 11/13$, leading naturally to a picture of
ordered di-vacancies. For $x = 0.71$, we find a hexagonal $
\sqrt{12}a \times \sqrt{12}a$ superlattice which contains 12 unit
cells.  Due to the unique triangular cobalt lattice structure and
its stacking sequence,
we find a special stability for a model of alternating
layers of Na tri-vacancy and quadri-vacancy, so that $x = 0.71$
is understood to be the average of 3/4 and 2/3 doping.
These results are supported by $^{23}$Na NMR.  Our analysis builds on
the vacancy clustering
model
suggested by Roger {\em et
al.}\cite{Roger}
However our finding that
tri- and
quadri-vacancies are more stable at smaller $x$ than
di-vacancies is contrary to their proposal.

Our picture of the Na vacancy leads naturally to an electronic model where a fraction of
Co${^{4+}}$ holes are bound to Na vacancies, resulting in the
co-existence of $S = {1\over 2}$ local moments and itinerant
carriers. For $x = 0.84$, we find an extremely low Fermi
temperature of 87~K for the itinerant carriers which may explain the
high thermopower and the emergence of ferromagnetism in the layer.
For $x = 0.71$, we suggest that the Curie-Weiss metal
may be indicative of a novel
quantum
spin liquid phase for the local
moments.

Na ions are easily lost during high temperature preparation,\cite{Motohashi} and they continue to diffuse to the surface layer even after stoichiometric sample is fully reacted and stored at room temperature.\cite{Shu}  Adding extra Na into the precursor has been a common practice in the powder sample preparation in order to compensate for the high temperature Na loss.  Alternatively ``rapid heat-up" technique is applied to reduce Na loss.\cite{Motohashi} However
neither
method
can
control Na level to the nominal concentration perfectly within a few percent error while 3-4 \% difference would introduce a completely different vacancy ordering patterns.  Many inconsistent results reported between $x = 0.67 - 0.85$ before are due to inaccurate Na content or inconsistent ordered patterns.
Thanks to
the much lower surface-to-bulk ratio in single crystal form, we believe structure work based on single crystal prepared using room temperature de-intercalation method is the most reliable route to study Na (vacancy) ordering.  The Na concentration of single crystal sample in this study has been checked thoroughly using Electron Microprobe Analysis (EPMA) from more than 10 points from freshly cleaved surface and c-axis parameters are calibrated against to those reported in the literature.  Combining our series of single crystal samples calibrated with EPMA\cite{Shu} and the reported powder sample calibrated with ICP analysis,\cite{Foo} the c-axis versus Na content relationship falls into a linear relationship of $c = -0.9866x + 11.639$ in the range of $ 0.6 \leq x \leq 0.84$.
As a byproduct of the present paper, we can recommend with confidence the use of this $c$-axis formula as a way of determining $x$.  This scale is anchored at two special points,
$x = 0.71$ and $x = 0.84$
where the $c$ axes
are $10.939{\AA}$ and $10.810{\AA}$, respectively.
As we shall see, based on the superlattice structure as supported by NMR, we arrive at these $x$ values independent of EPMA analysis.

\begin{figure}
\includegraphics [width=3.3in]{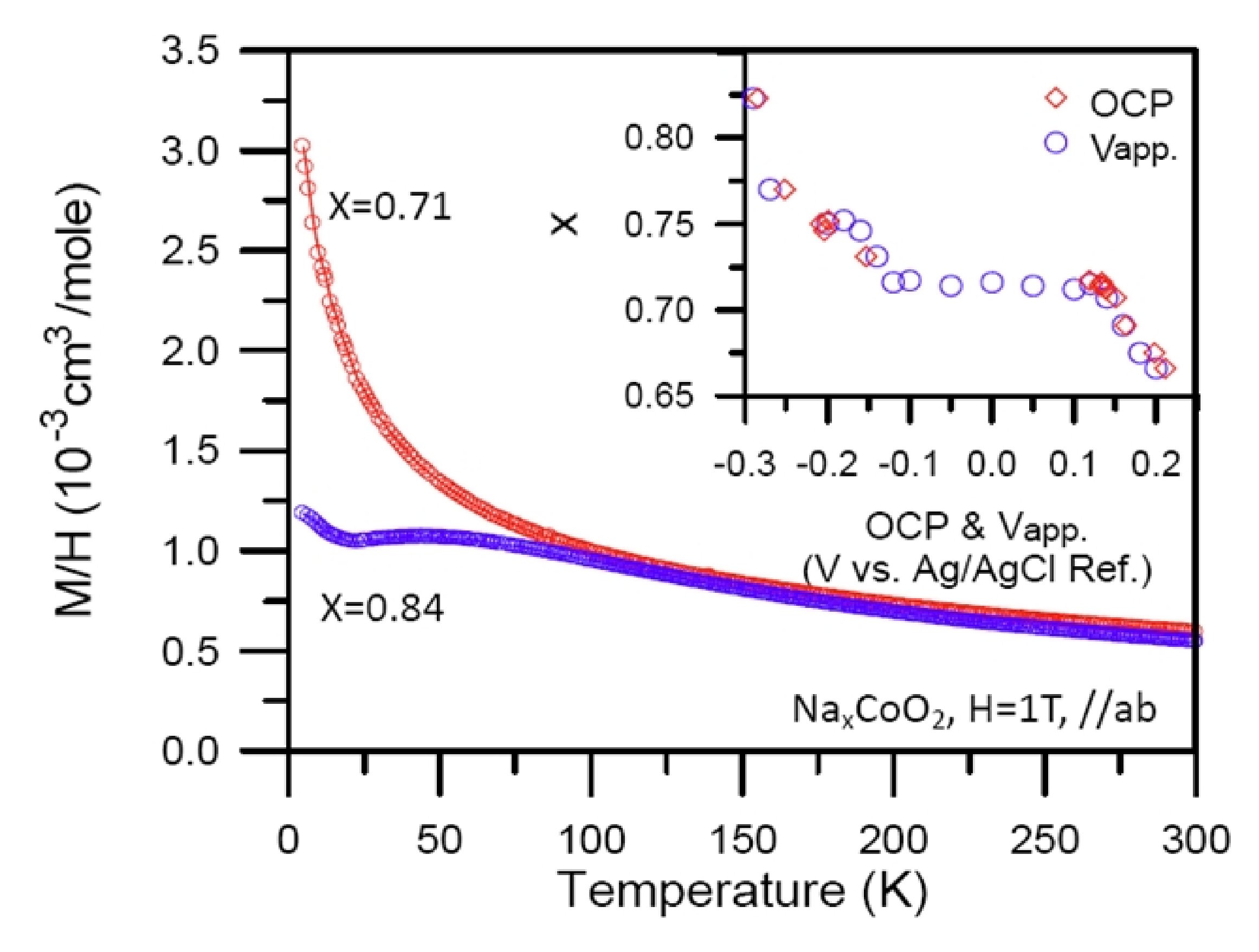}
\caption{\label{fig:Fig-1} {\bf Spin susceptibilities and preparation method.}  The spin susceptibility for field parallel to the ab plane for single crystal samples of $x = 0.71$ and $0.84$.  Note the strong increase for $x = 0.71$ at low temperatures.  The inset shows the open circuit potential (OCP) and the applied potential $V_{\rm app}$ in our electrochemical cell.  The step at $x = 0.71$ indicates a concentration of special stability.}
\vspace{-5mm}
\end{figure}

\vspace*{.25in} $\gamma$ phase $Na_xCoO_2$ is prepared by
chronoamperemetry technique on single crystal sample in
electrochemical cell constructed as $Na_xCoO_2$/1N $NaClO_4$ in propylene carbonate/Pt.\cite{Shu}
Starting from the original crystal $Na_{0.84}CoO_2$ grown with
floating-zone method, constant anodic potential is applied to the
sample electrode until the induced current decays to zero. The
concentration $x$ has been carefully determined by freshly cleaved
surface using EPMA. The applied
voltage (V$_{app}$) for prepration and its final equilibrated open
circuit potential (OCP) versus $x$ is summarized in Fig.\ref{fig:Fig-1}.
The most prominent feature in Fig.\ref{fig:Fig-1} is the step at $x = 0.71$
vs applied voltage, indicating that $x = 0.71$ is an exceptionally
stable phase. This is very surprising because the naive expectation
is for steps at rational fractions such as ${3 \over 4}$ or ${2
\over 3}$, but such steps are either weak or non-existent.

Another indication that $x = 0.71$ is a specially ordered structure comes from the fact that crystals of exceptional homogeneity can be grown at this concentration.  The residual resistivity of $x = 0.71$ is low among the cobaltate family\cite{Foo} and has been measured by M. Lee and N. P. Ong to be as low as $8 \mu\Omega$-cm at 0.3~K in our crystals.  This is seven times lower than commonly reported in the literature\cite{Li} and corresponds to $k_F\ell = 220$ where $\ell$ is the mean free path.  The special order also leads to narrow NMR lines which we will take advantage of later.

Another special
concentration is $x = 0.84$, which is the as-grown material obtained
using optical floating-zone technique.\cite{Chou} For $x$ greater
than 0.84 we generally find mixed phases, in agreement with the
observation by Lee {\em et al}.\cite{Lee} Between $x = 0.75$ and
0.84 we find single phase with a magnetic phase transition around
20~K. At $x = 0.84$ the transition moves up to 27~K. The magnetic order is known to be A type, with ferromagnetic
order in plane and antiferromagnetic between neighboring Co
planes.\cite{Bayrakci05} This has presented a puzzle, because the
spin susceptibility for $x = 0.84$ powder average above 100~K (see Fig.\ref{fig:Fig-1}) shows a Curie-Weiss fit $\chi = \chi_0 + C/(T-\theta)$ with $\theta \approx
-77$ K, characteristic of antiferromagnetic exchange.  At $x = 0.71$
the magnetic order disappears. The spin susceptibility above 100~K
is very similar to that of 0.84, but keeps rising at low
temperatures. Data below 60~K can be fitted by a Curie-Weiss law
with $C \approx 0.04 ~cm^3~K/mole$ and $\theta \approx -12$ K. This
rise is much stronger than previously reported. \cite{Foo}  The
resistivity shows a linear T behavior below 100~K and becomes T$^2$
only below $\approx 4$ K.\cite{Li} The name ``Curie-Weiss'' metal
has been given to this highly unusual combination of magnetic and
transport properties.\cite{Foo}  Due to their special properties and
stability, we focus our attention on $x = 0.84$ and $0.71$.

\begin{figure}
\includegraphics [width=3.3in]{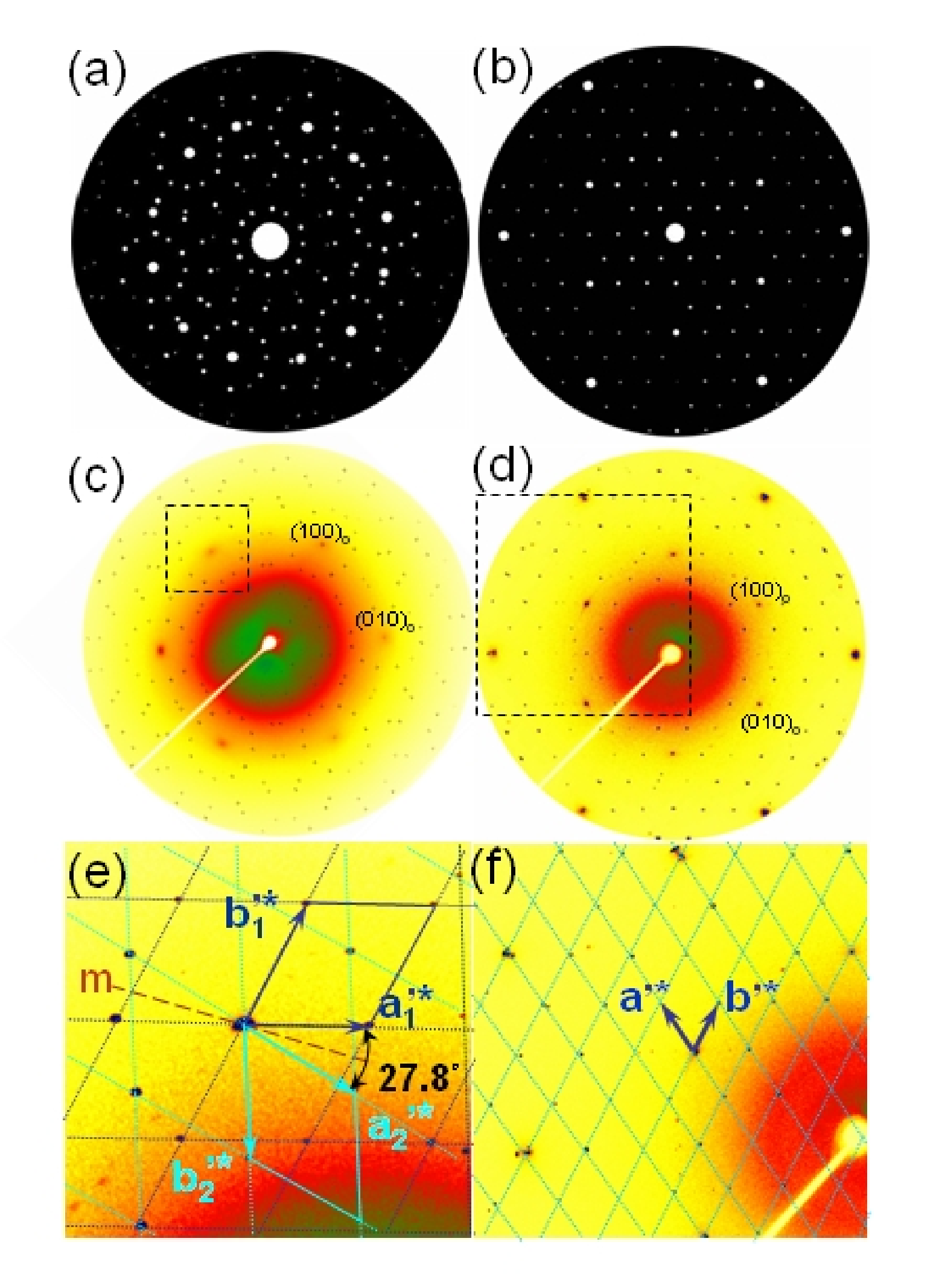}
\caption{\label{fig:Fig-2} {\bf Single crystal synchrotron X-ray diffraction and analysis} The Laue patterns for (c) $x = 0.84$ and (d) $x = 0.71$ and the corresponding simulated diffraction patterns (a) and (b).  Note the 12 member ring for $x = 0.84$ in (c) and the simple hexagonal superstructure for $x = 0.71$ in (d).  (e) is a blow up of the dashed region in (c) showing that the superstructure for $x = 0.84$ can be indexed with two sets of $\mathbf{a^*}$ in k-space ($\sqrt{13}a$ in real space) of 27.8$^\circ$ separation with a mirror plane. (f) is a blow up of the dashed region in (d) showing single domain indexed for $x = 0.71$, which corresponds to a hexagonal superlattice of $\sqrt{12}a$ in real space.}
\vspace{-5mm}
\end{figure}

\begin{figure}
\includegraphics [width=3.3in]{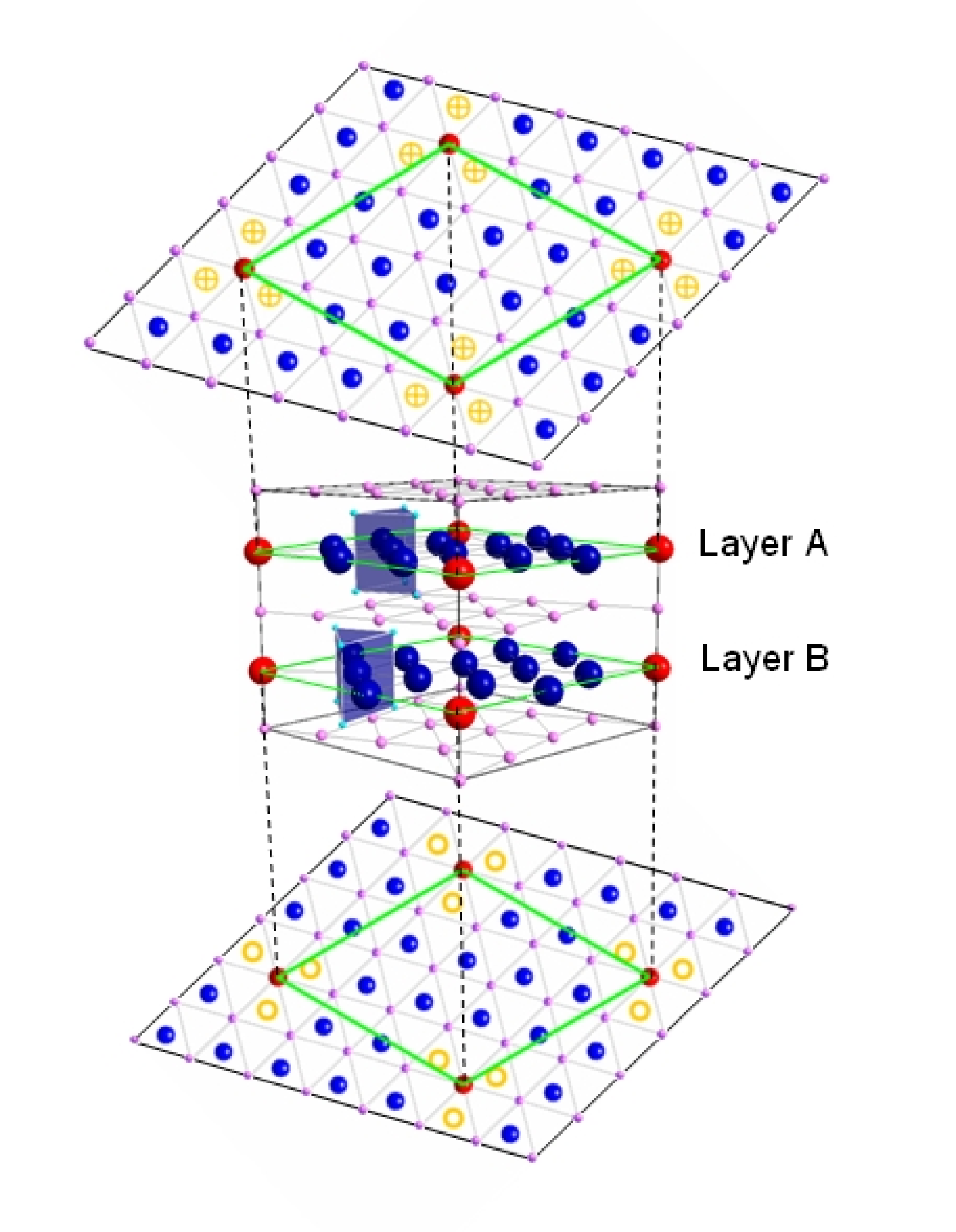}
\caption{\label{fig:Fig-3} {\bf 2d and 3d structure model of $Na_{0.84}CoO_2$.} The middle panel shows the unit cell of $Na_{0.84}CoO_2$, where oxygen atoms are removed for better viewing, except for the prismatic cage near Na2.  The Na2 site (blue) sits at the center of an oxygen prism (shaded blue) which is shifted between the two layers of the unit cell.  The Na1 site (red) sits directly above and below Co (light blue).  Top and bottom panels show the 2D view of the top(A) and bottom (B) Na layers.  The green lines show the $\sqrt{13}a \times \sqrt{13}a$ superlattice formed by the di-vacancy which is centered on the Na1 site surrounded by 3 empty Na2 sites (yellow).}
\vspace{-5mm}
\end{figure}

We performed transmission Laue X-ray diffraction study on single
crystal $Na_xCoO_2$ with x = 0.84 and 0.71 using synchrotron
source of Taiwan NSRRC.  Superlattice diffraction spots appear near
the original $P6_3/mmc$ structure indices \{100\}, \{110\} and \{200\} as
shown in Fig.\ref{fig:Fig-2}.  For $x = 0.84$ we found a hexagonal
${\sqrt{13}a \times \sqrt{13}a}$ superlattice structure.
The major character of this Laue pattern is the
12 spots ring that formed near \{100\} peaks of the original lattice.
As shown in the Fig.\ref{fig:Fig-2}(c),
these spots consist of 2 sets of six-fold rings rotated by $\sim 28^\circ$, where each set is indexed using two simple hexagonal unit cells with $a^\prime = \sqrt{13}a$.  Direct evidence of this model comes from electron diffraction with a beam size of $150-200$~nm, which shows 6-fold rings coming from one of two domains.
Details are discussed in the supplementary section.  Fig.\ref{fig:Fig-2}(a) shows the simulated diffraction pattern along the c-axis for $x = 0.84$, where the agreement using two-domain simulation is high even before the intensity is corrected by the refined Na2 position and the associated structure factors.
The ${\sqrt{13}a \times \sqrt{13}a}$ structure corresponds precisely to
the ordering of di-vacancies shown in Fig.\ref{fig:Fig-3}.  With one
di-vacancy per 13 cobalt, we predict $x = 1 - {2\over 13} = 0.846$
which agrees with
$x = 0.84 \pm 0.01$ measured with EPMA within error.

We note that Roger {\em et al}.\cite{Roger} reported very similar 12 spot ring diffraction patterns for what they labeled as $x = 0.75$.  However, they fitted the pattern with a 15-site monoclinic superstructure which contains one tri-vacancy.  This model gives a Na content of $x = 1-{3 \over 15} = 0.8$, inconsistent with $x = 0.75$.  We suspect that their sample grown by high temperature quench may suffer from inhomogeneity and uncertainty with respect to $x$ determination.  It will be useful to check if their pattern can also be indexed with our structure model.

Next we discuss the
stacking properties
of the Na vacancies.
We first review the
salient feature of the parent $\gamma-Na_xCoO_2$ lattice of $P6_3/mmc$
symmetry using $x = 0.84$ unit cell as shown in Fig.\ref{fig:Fig-3}. The unit cell consists of two layers of cobalt atoms located directly on top of each other and layers of Na atoms sandwiched in between. There are two Na sites in each layer, called Na1 at (0,0,$\frac{1}{4}$) and Na2 ($\frac{2}{3}$,$\frac{1}{3}$,$\frac{1}{4}$), the latter
being the preferred site because Na1 sits directly on top of the
positively charged Co and is more costly electrostatically.  Na2
sits in the center of a prismatic cage formed by six oxygens in each layer, and the two Na layers (labeled $A,B$) are distinguished by the different position of these prismatic
cages.  In addition to the Na sites, there are sites sitting directly under oxygen which we refer to as O sites.  Note
that in layer A
the Na2 and O sites occupy the center of the down and up pointing triangles, respectively, but their positions are interchanged when we go from
layer A to layer B.  This will have crucial consequences later.  The di-vacancy is formed by creating three Na2 vacancies (shown in yellow) and then
occupying a Na1 site.\cite{Roger}  Since the di-vacancies are
centered on the Na1 site, they can stack directly on top of each
other, going from layer A to B.  On the other hand, out of phase
stacking, for example by placing the di-vacancy at the center of the
green diamond shown in Fig.\ref{fig:Fig-3} produces ambiguity because two
such sites are equivalent. This will lead to disorder.  At present
it is not possible to distinguish between in phase and out of phase
stacking by refinement.

\begin{figure}
\includegraphics [width=3.3in]{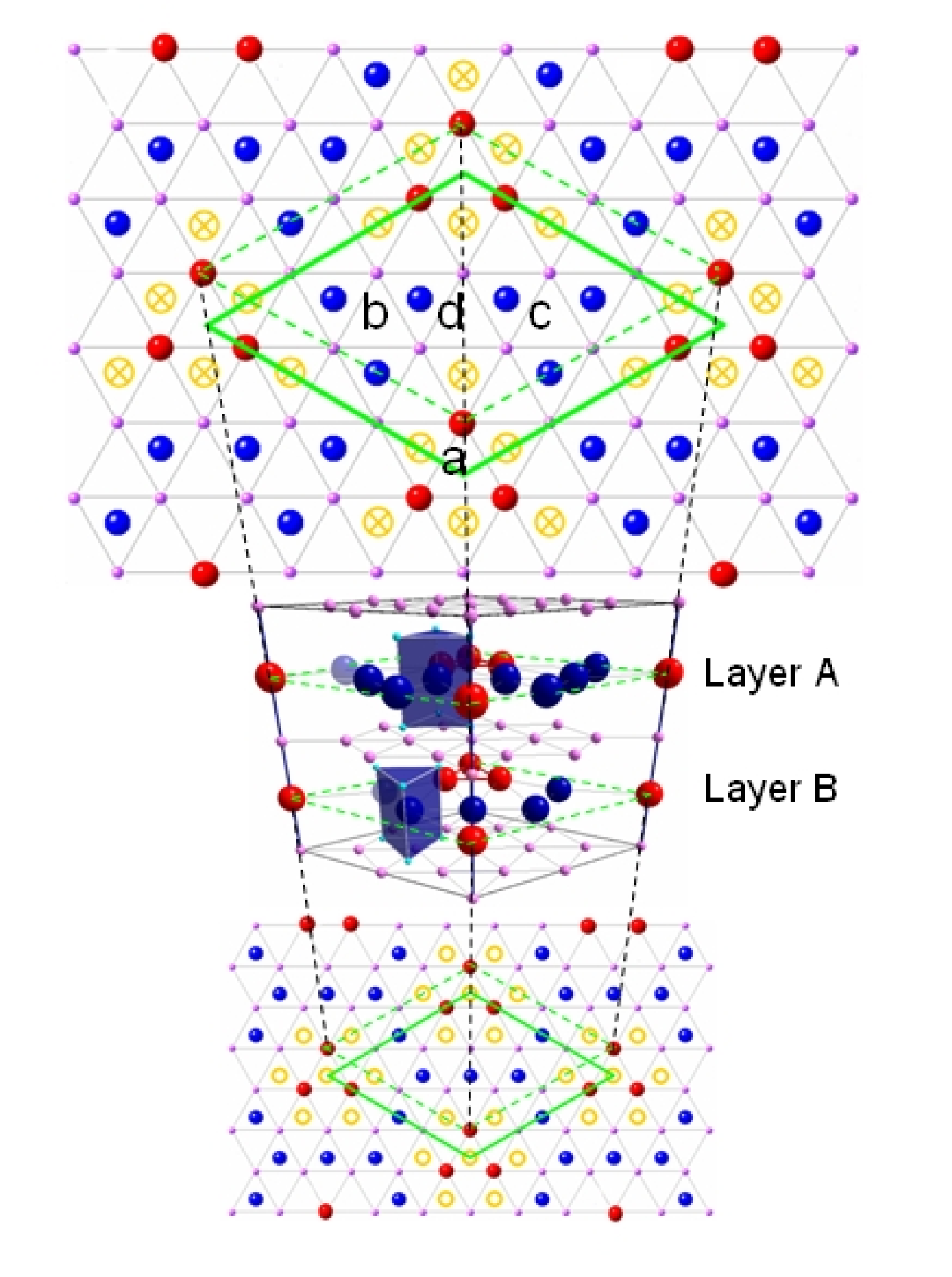}
\caption{\label{fig:Fig-4} {\bf 2d and 3d structure model of $Na_{0.71}CoO_2$.} The unit cell for x = 0.71 before considering the 3c stacking.  The $\sqrt{12}a$ superlattice structure of tri-vacancy in layer A and quadri-vacancy in layer B are shown with green solid lines.  Note the vacancies consist of Na1 trimers surrounded by empty Na2 sites (yellow circles).}
\vspace{-5mm}
\end{figure}

Next we turn to the $x = 0.71$ sample
which,
as mentioned earlier,
is special in its stability.
Density Function Theory
(DFT) model calculations indicate special stability for $x = 5/7 =
0.714$ and our initial expectation was to search for hexagonal ${\sqrt{7}a
\times \sqrt{7}a}$ superstructure of di-vacancies.\cite{Zhang}
Instead, we discovered hexagonal superstructure at ${\sqrt{12}a \times
\sqrt{12}a}$ which contains 12 Na2 sites per unit cell per layer. We
are then led to consider tri- and quadri-vacancies in order to
account for the 0.29 missing Na ions.  These vacancies were
described by Roger {\em et al}.\cite{Roger} and shown in Fig.\ref{fig:Fig-4}, where
we emphasize the special role of A,B stacking.  The tri- and
quadri-vacancies consist of Na1 trimers surrounded by six and seven
Na2 vacancies (shown in yellow), respectively. Note that the trimer
in the tri-vacancy is centered on the O site while the trimer in the
quadri-vacancy is centered on the Na2 site.  Recall that the role of
O and Na2 sites are reversed between layers A and B.  This has the
remarkable consequence that tri-and quadri-vacancies can stack
coherently, e.g. they can either (i)
stack in phase, with tri-vacancy and quadri-vacancy sitting
directly on top of each
other, or (ii)
stack out of phase, with
tri-vacancy on site a in layer A and quadri-vacancy either
directly below the
d site or
below
the b,c sites. (these sites are defined in Fig \ref{fig:Fig-4}.)
The alternating stacking of tri- and quadri-vacancies lead to $x = 1 - {1\over 2} \left( {3+4\over 12} \right) = 0.708$. We believe the special
stacking property described above is responsible for the exceptional
stability of the 0.71 Na concentration.

We expect out of phase stacking to be more favorable by electrostatic consideration.
In this case the  c-axis periodicity could be multiplied due to a versatile AB stacking combinations, e.g. A$_a$B$_b$ (c), A$_a$B$_b$A$_a$B$_c$ (2c) or A$_a$B$_b$A$_c$B$_a$A$_b$B$_c$ (3c).  We have found evidence of 3c modulation through both single crystal electron diffraction as well as synchrotron X-ray powder diffraction techniques, which is discussed in detail in the supplementary section.  The evidence of 3c modulation
supports A$_a$B$_b$A$_c$B$_a$A$_b$B$_c$ stacking and
indicates that simple electrostatic consideration
dominates the stacking sequence of vacancy cluster units between layers.

We next consider stacking of tri-vacancies in the AB layers (TT)
to form
$x = 1 -
\frac{1}{2}(\frac{3+3}{12})= 0.75$, or
stacking quadri-vacancies
(QQ) to form
$x = 1 - \frac{1}{2}(\frac{4+4}{12})= 0.67$.
Note that unlike the 0.71 case considered earlier,
there are now three equivalent choices to stack the center of Na1 trimers in the case of TT and QQ stacking.
This is true for both in-phase and
out-of-phase stacking.
Such triangular ambiguity introduces significant disordering for the TT and QQ vacancy clusters, which could be the major reason why $x = 0.71$ is
the most stable phase among these three types of stacking. Indeed,as shown in Fig.\ref{fig:Fig-1}, the stability range for $x = 0.75$ and 0.67 are relatively narrow as indicated by the $dx/V_{app}= 0$ plateau in the electrochemical preparation.

\begin{figure}
\includegraphics [width=3.3in]{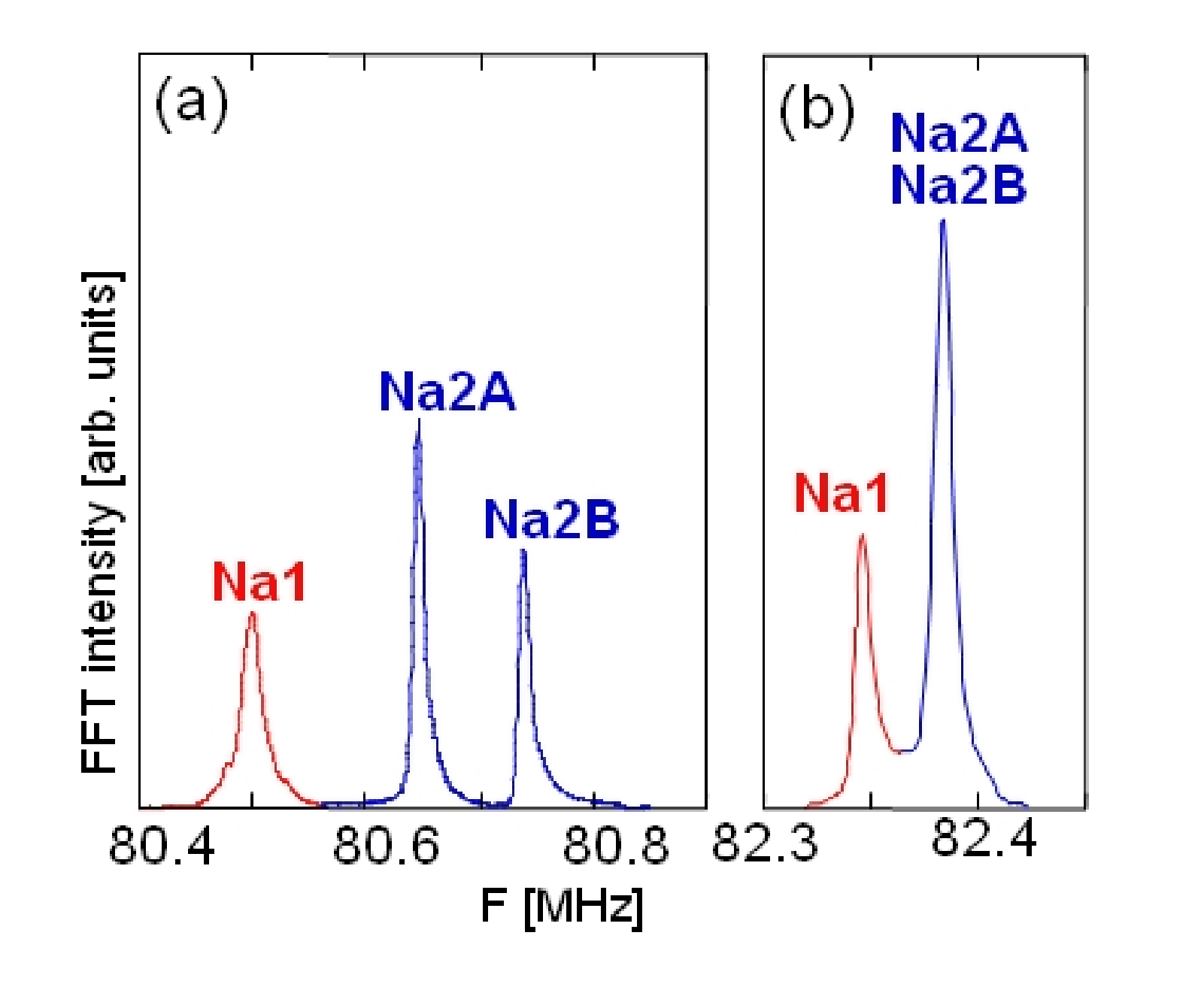}
\caption{\label{fig:Fig-5} {\bf $^{23}$Na NMR spectra for $Na_{0.71}CoO_2$.} $^{23}$Na NMR spectra for $x = 0.71$ taken at 20~K.  Applied field B = 7.298~Tesla is parallel to the c-axis.  (a) the $\pm {3\over 2}$ to $\pm {1\over 2}$ satellite transitions, and (b) the $\left( -{1\over 2} \rightarrow {1\over 2} \right)$ central transitions. } \vspace{-5mm} \end{figure}

We have also performed $^{23}$Na NMR measurements of $x = 0.71$ crystal.  The central transition $\left(  {1\over 2} \rightarrow -{1\over 2} \right)$ and the lower satellites $\left(  \pm {3\over 2} \rightarrow \pm {1\over 2} \right)$ are shown in Fig.~5.  These lines are considerably narrower than what were previously reported in the literature,\cite{Mukhamedshin,Ning} and are also narrow compared with our crystals at other Na concentrations.  The central line in Fig. 5(b) separates into two groups with different Knight shifts and we assign the minority as
Na1 as indicated in the figure.  The satellite peaks of Na2 further split into two peaks, Na2A and Na2B, due to the difference in the nuclear quadrupole interaction.  We measured the weight of the Na1~:~Na2A~:~Na2B lines to be 36.3\%~:~37.1\%~:~26.5\% with an error of $\pm 3\%$.  In our structure model we have three Na1 and six Na2 in layer A (tri-vacancy layer) and three Na1 and five Na2 in layer B (quadri-vacancy layer).  The Na1 form trimers and have similar environments in both layers. The total fraction of Na1 is predicted to be $\frac{3+3}{3+3+6+5} = 35.3\%$, in excellent agreement with observation.  We emphasize that the sharpness of the lines allows accurate weight determination which can be used to rule out other structure models.  For example, if layer B were to contain two di-vacancies instead of a single quadri-vacancy per unit cell, leaving the $x$ value unchanged, that layer would have two Na1 and six Na2 sites and the total fraction of Na1 is predicted to be $\frac{3+2}{3+2+6+6} = 29.4\%$, inconsistent with the measured value.
Generally, the existence of di-vacancy and mono-vacancy will decrease the Na1 fraction beyond what is permissible, giving us great confidence in our assignment of tri- and quadri-vacancy to $x = 0.71$.  We also note that the six Na2 sites have identical environments in layer A, while in layer B five Na2 are separated into two groups of 2 sites with three Na2 nearest neighbors and 3 sites with two Na2 and one Na1 nearest neighbors.  Provided that the splitting of the latter group is not resolved, it is tempting to assign the Na2A and Na2B lines to the A and B layers, respectively, in which case the weight ratio of Na1 : Na2A : Na2B is predicted to be 6~:~6:~5~=~35.3\%~:~35.3\%~:~29.4\%, within error of the observation.

What are the consequences of the Na superstructure order for the
electronic properties of the cobalt oxide planes?  First we consider
$x = 0.84$. Starting from the band insulator $NaCoO_2$,  there are
2/13 holes in an otherwise filled band.
It is natural that one hole will be bound to each
di-vacancy, forming a localized $S = {1\over 2}$ spin.  The second
hole remains itinerant, explaining the metallic nature of this
state.  This leads to
a model
of local moments
on a triangular lattice
interacting with an antiferromagnetic Heisenberg exchange $J_H$
and
coupled to
an equal density of conduction
electrons via a Kondo coupling $J_K$.
This model
is not sufficient to describe the physics of $x = 0.84$
cobaltate, because the ground state is predicted to be either a
metal with an antiferromagnetically ordered moment (with nearest
neighbor spins forming a $120^\circ$ angle) or a Kondo insulator,
whereas the experiment requires a metallic ground state with
ferromagnetic order in the plane.  To see where the ferromagnetism
comes from, we need to put in some numbers.  The specific heat shows
a peak at the magnetic transition $T_N$ with a low temperature
linear $T$ coefficient $\gamma$ of $\approx
10$~mJ/Co-mole.\cite{Bayrakci04} The extrapolation of the linear $T$
contribution from above $T_N$ gives $\gamma \approx 24$
mJ/Co-mole.\cite{Luo}  Assuming a spin unpolarized parabolic band,
the latter value gives a mass $m^\ast = 35$ m$_e$.  In a nearest
neighbor tight binding fit to the band structure, this corresponds
to a hopping matrix element $t_{eff} \approx 14$ meV.  This is a
factor of 6 or 7 mass enhancement compared with the typical tight
binding fit to LDA bands.  Coulomb repulsion is presumably
responsible for this enhancement.  Assuming this $m^\ast$ and a hole
density of ${1\over 13}$ carriers per Co, we obtain a Fermi
temperature of 87 K.  This exceptionally low energy scale has
several important consequences.

First, in a narrow band with low energy, we expect a ferromagnetic
tendency due to the Stoner mechanism, i.e., it is favorable to spin
polarize to gain exchange energy at the expense of kinetic energy.
Indeed, we believe that the ground state is fully spin
polarized to form a ``half-metal.''  This will explain why the
$\gamma$ term in the specific heat drops by approximately a factor
of 2 (from 24 to 10 mJ/Co-mole) across $T_N$.  Furthermore, the {\em
fully polarized} band has a Fermi surface area exactly equal to that
of the reduced Brillouin zone (both equal to ${1\over 13}$ of the
full zone).  Intersection of an almost circular hole Fermi surface
with a hexagonal reduced Brillouin zone (both equal to ${1\over 13}$
of the full zone)  gives rise to small hole pockets and electron
pockets, the latter being centered at the zone corner.
Recently, L. Balicas and co-workers have observed small pockets in quantum oscillation studies of our x=0.84 crystals, which lends support to this picture.

The physical picture is that the ferromagnetic tendency among the
conduction electrons opposes the Kondo coupling to form a singlet
with the local moments, thereby suppressing the Kondo insulator
state. The $S = {1\over 2}$ Heisenberg model on a triangular lattice
is subject to strong quantum fluctuations and the $120^\circ$ order
may be greatly suppressed or eliminated entirely due to the Kondo
coupling. The full many-body problem is a complicated one which
deserves a separate study.  In the supplementary information section we outline a simple mean field
treatment
which captures some features of the susceptibility.

Here we emphasize that the very small Fermi temperature helps explain the enhanced
thermopower for $x = 0.84$.  Conventional Sommerfeld theory predicts
$S = (k_B/e) T/T_F$ for $T < T_F$ and $S$ is expected to roll over
above $T_F$ to a constant which depends on the classical
configuation (the Heikes formula).\cite{Chaikin} Experimentally, $S$
shows a shallow peak around 100~K for $x = 0.84$ and the fit to low
temperature Sommerfeld theory yields $T_F \approx$ 30~K.  We see that  the
crossover scale is about right and explains the success of the
Heikes formula at these relatively low temperatures,\cite{Mukerjee}
while the overall linear $T$ slope and the magnitude need a factor of 3 enhancement compared
with our estimate of $T_F \approx 87$~K.  One possible source of
enhancement is the deviation from a parabolic band because the band
calculation shows severe flattening and even a dimple at the
$\Gamma$ point.  Thermopower relies on the breaking of particle-hole
symmetry, which is enhanced by a mass which decreases with
increasing hole energy. Supporting evidence of this possibility
comes from the observation that the low temperature $\gamma$ term
for the spin polarized band (10~mJ) is less than half of that
extrapolated from high temperature (24~mJ).  A second indication of
the small Fermi energy comes from the Hall effect.  The Hall
constant $R_H$ is predicted to be linear in $T > T_F$, a special
consequence of hopping on a triangular lattice.\cite{Motrunich}
Experimentally $R_H$ is linear in $T$ above $\sim 100$~K, and
saturates to a very small value at low temperature.\cite{Foo}  The
latter may be a consequence of cancellations between the small
electron and hole pocket mentioned earlier.

Optical measurements provide direct evidence for heavy mass and
small carrier density.  The low frequency conductivity consists of a
broad peak centered around 150~cm$^{-1}$ and a narrow Drude peak
with a spectral weight estimated to be ${1\over 6}$ of that of the
broad peak.\cite{Bernhard,Wang}  We associate the broad peak with
the hole bound to the Na vacancy.  The Drude weight was estimated to
correspond to $\omega_{\rm plasma} = \left( 4\pi ne^2/m
\right)^{1\over 2} \approx 1300$~cm$^{-1}$ from which $n/m$ can be
extracted.\cite{Bernhard}  If we assume $m^* = 35 m_e$, we find $n$ to
be even smaller $\left( {\rm about} {1\over 3} \right)$ than our
estimate based on our model of 1/13 mobile carrriers per Co. This
extremely small Drude weight is very difficult to understand if all
the holes are mobile.

Next we turn our attention to $x = 0.71$.  This state, dubbed the
``Curie-Weiss metal'' is in many ways the most mysterious among all
the doping concentrations.  The $T_N$ transition disappears and
$\chi$ keeps rising with decreasing $T$.  The resistivity $\rho (T)$
is linear down to $\sim 4$~K.  The uniqueness of this behavior even
in comparison with heavy fermion compounds has been emphasized by Li
{\em et al}.\cite{Li}

The observation of
Curie-Weiss behavior in a metal again calls for the coexistence of local
moments and itinerant electrons.  As an example, let us assume out
of phase stacking of tri- and quadri- vacancies and that the
quadri-vacancy binds a hole on each of the neighboring Co planes,
forming an $S={1\over 2}$ local moment on each plane.  The local
moment forms a triangular lattice with one $S = {1\over 2}$ per 12
cobalt.
For $x = 0.71$, the
concentration of itinerant carriers is 2.5/12 $\approx$ 0.208 per
Co, which is 2.7 times that of $x = 0.84$.  We then expect a Fermi
temperature of 235~K.  The higher carrier density reduces the
tendency towards Stoner instability, which explains the absence of a
magnetic phase transition.  In support of this picture, we note that
in the thermopower, the saturation scale has moved to room
temperature or above\cite{Lee} and the fit of the low $T$ data to
$(k_B/e) T/T_F$ yields $T_F \approx 200$~K, consistent with our
estimate and much larger than that for $x = 0.84$.

Finally, we note that a picture of conduction electrons coexisting
with local moments which do not order down to the lowest temperature
suggests the existence of
a novel ground state for the spin system called the quantum spin
liquid.  While this possibility has been
proposed
in the
literature,\cite{Senthil} there has been no known experimental
realization.  A system of $S = {1\over 2}$ on a triangular lattice
coupled to a narrow band of conduction electrons with some
ferromagnetic tendency may provide just the needed frustration to
stabilize such a state.  We note that spin liquid on a triangular
lattice is expected to support a spinon Fermi surface,\cite{S-SLee}
with $\chi^{\prime\prime}(q,\omega)$ predicted to behave
as\cite{Altshuler}

\begin{equation} \label{chi}
\chi^{\prime\prime} (q,\omega) \approx \left[ \left| |q| - Q_0
\right|^{3\sigma-1} + \left( {\omega\over \omega_0}\right) ^{{2\over
3} (3\sigma-1)} \right]^{-1}
\end{equation}
where $Q_0$ is twice the spinon Fermi vector and $\sigma$ is a
critical exponent which was estimated to be 0.35 in a large $N$ expansion.
Equation ~(\ref{chi}) has unique signatures which can be
tested by neutron scattering.

In conclusion, the observation of Na superlattice structure in the
special Na concentrations of 0.84 and 0.71 show the importance of Na
ordering in determining the unusual electronic properties.  We have
introduced a minimal model of local moments coexisting with
itinerant electrons which accounts for many of the puzzling features
observed in $x = 0.84$ and which forms the starting point for
unraveling the mystery surrounding the ``Curie-Weiss metallic''
phase of $x = 0.71$.  We believe exciting discoveries of exotic
states of matter may lie ahead.





\end{document}